\documentclass{INTERSPEECH2023}
\interspeechcameraready 

\usepackage{ragged2e}
\usepackage[export]{adjustbox}
\usepackage{tabularx}
\usepackage{float}
\usepackage{multirow}
\usepackage{multicol}
\restylefloat{table}
\usepackage{cite}

\title{Robust and lightweight audio fingerprint for Automatic Content Recognition}
\name{Anoubhav Agarwaal, Prabhat Kanaujia, Sartaki Sinha Roy, Susmita Ghose}
\address{Data Science, LG Ads Solutions, Mountain View, California, USA}

\email{anoubhav@lgads.tv, prabhat@lgads.tv, sartaki@lgads.tv, susmita@lgads.tv}

\begin{document}

\maketitle

\begin{abstract}
\noindent 
This research paper presents a novel audio fingerprinting system for Automatic Content Recognition (ACR). By using signal processing techniques and statistical transformations, our proposed method generates compact fingerprints of audio segments that are robust to noise degradations present in real-world audio. The system is designed to be highly scalable, with the ability to identify thousands of hours of content using fingerprints generated from millions of TVs. The fingerprint’s high temporal correlation and utilization of existing GPU-compatible Approximate Nearest Neighbour (ANN) search algorithms make this possible. Furthermore, the fingerprint generation can run on low-power devices with limited compute, making it accessible to a wide range of applications. Experimental results show improvements in our proposed system compared to a min-hash based audio fingerprint on all evaluated metrics, including accuracy on proprietary ACR datasets, retrieval speed, memory usage, and robustness to various noises. For similar retrieval accuracy, our system is 30x faster and uses 6x fewer fingerprints than the min-hash method.

\end{abstract}

\section{Introduction}
Audio fingerprinting systems have emerged as a powerful tool for identifying audio content without requiring access to the actual audio signal. One of the primary applications of audio fingerprinting is Automatic Content Recognition (ACR). ACR involves the identification of various forms of audio content, such as songs, TV shows, and movies. The technology works by converting the audio being played on a device into fingerprints, which are subsequently matched against a database. \cite{Cano}. Other applications include music retrieval \cite{Gfeller, Wang, Six, dejavu}, watermarking, and copyright detection \cite{Cano, Gomez}, content de-duplication \cite{Burges}, and broadcast monitoring \cite{Allamanche}. 

Fingerprints are a compact representation of high sample rate raw audio data, which retain the key information required to uniquely identify a given audio segment. Using fingerprints for matching instead of raw audio offers several benefits. One advantage is reduced memory consumption for the query and the reference database, which it is matched against. Additionally, it lowers the bandwidth requirements for transmitting the query fingerprints from the device. By using fingerprints, there is also a reduction in computational demands for content identification while searching in the database. Furthermore, the use of fingerprints enables systems that are robust to noises and degradations \cite{Haitsma}. 

Our proposed approach retains the advantages of a typical audio fingerprint while also meeting the specific requirements of running ACR on TV devices.

Firstly, our solution should be highly scalable to identify contents using fingerprints generated from millions of devices. Our approach is designed to have a high temporal correlation, resulting in fingerprints from highly overlapping audio regions being similar in terms of distance. This allows for the creation of sparse reference databases without any significant loss of retrieval accuracy, enabling much faster retrieval speeds. Furthermore, our proposed fingerprint is represented in the Euclidean or L2 space, and it can leverage existing GPU-compatible ANN search algorithms.

Secondly, our solution should be lightweight, given the limited computational resources available on low-power TVs running ACR. As such, the type of transformations that can be performed to convert an audio representation to a lower dimensional fingerprint has been restricted. Initially, we experimented with neural-network-based fingerprints inspired by \cite{Sungkyun, Gfeller, Purwins}. However, they were computationally expensive to generate, and the memory consumption was an order of magnitude more than what is able on the TVs. Thus, signal processing techniques and simpler statistical transformations were used to arrive at our proposed fingerprint.

The paper is organized as follows: In Section \ref{sec:audio_fp}, we describe the algorithm used to obtain this novel audio fingerprint (Figure~\ref{fig:2}). Our approach involves several transformations, including time averaging, standardization, amplitude deltas, and principal component analysis, which are discussed in detail. These transformations aim to extract key features from the audio signal while reducing the dimensionality of the fingerprint, thus further reducing its memory footprint. Section \ref{sec:exp_method} describes the experimental methodology. Section \ref{sec:exp_results} presents the results for our proposed fingerprinting approach, including its robustness to noise, retrieval speed, and temporal correlation and sparsity. We compare our approach with a conventional min-hash based audio fingerprint \cite{Baluja}, and we test its performance on industry-scale proprietary datasets used for ACR on a TV and standard artificial degradations commonly used in fingerprinting literature \cite{Haitsma}. Finally, we conclude the paper with Section \ref{sec:conc}, which summarizes our contributions and discusses future work.

\begin{figure*}[!htbp]
    \centering
    \includegraphics[width=\textwidth]{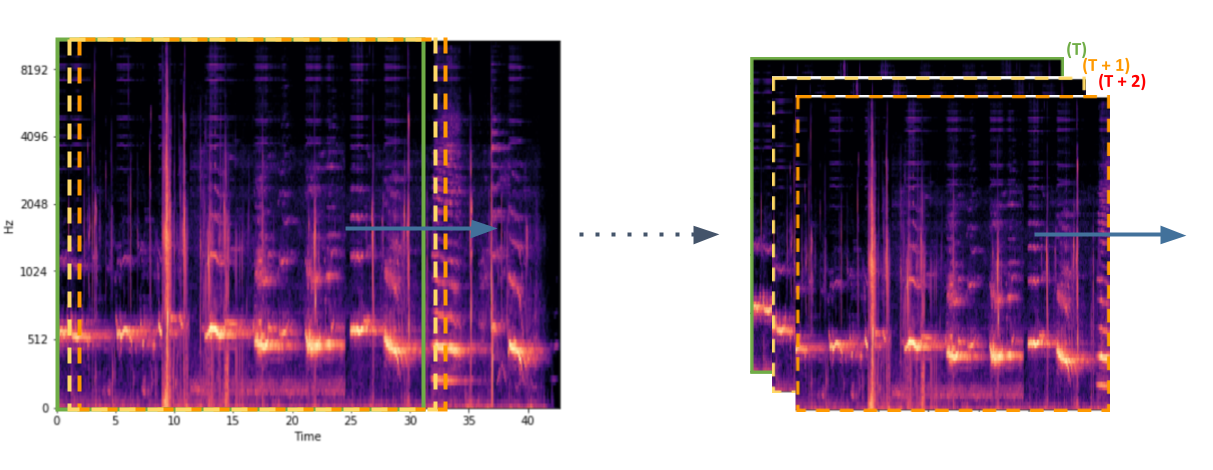}
    \caption{Windowing operation in section~\ref{sec:window}}
    \label{fig:1}
\end{figure*}

\section{Proposed fingerprint generation}
\label{sec:audio_fp}
\begin{figure*}
    \centering
    \includegraphics[width=\textwidth]{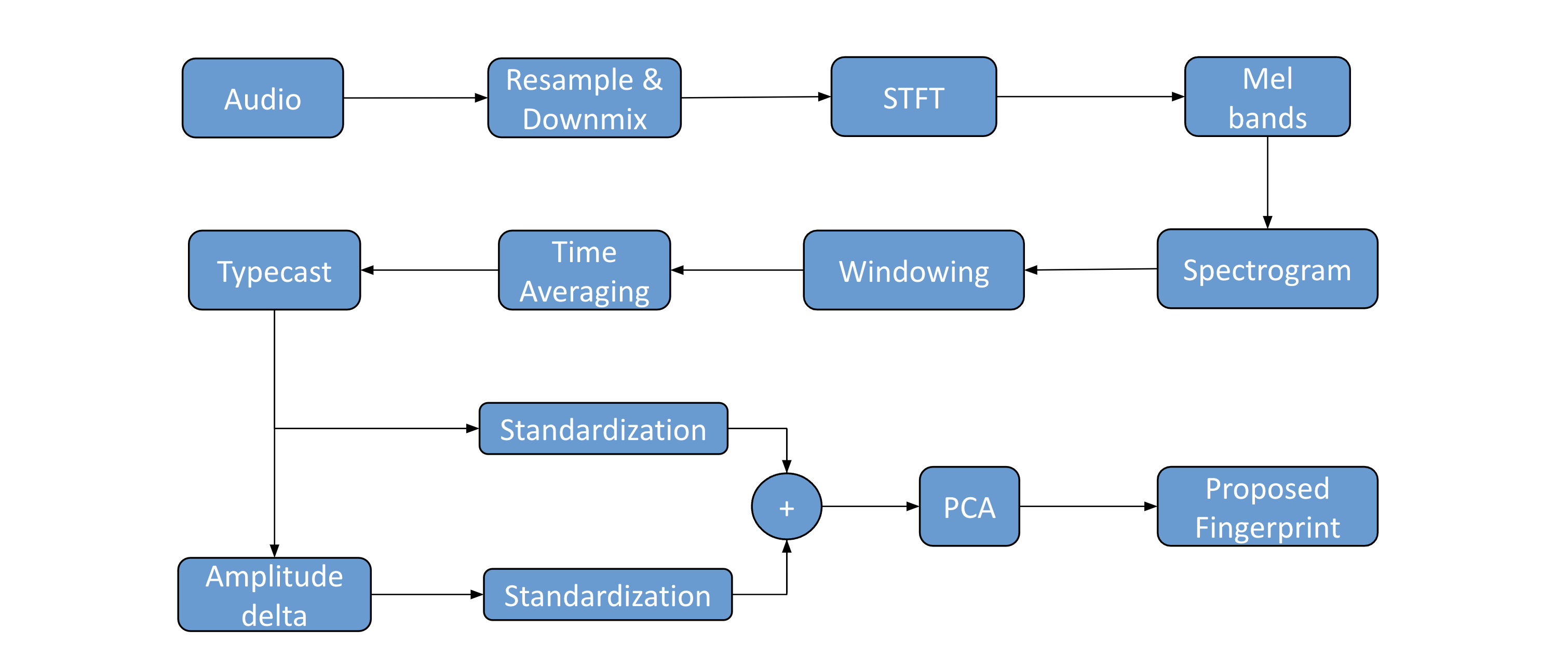}
    \caption{Proposed fingerprinting pipeline}
    \label{fig:2}
\end{figure*}

The components of our proposed fingerprinting algorithm are explained in the subsequent sections and visualized in Figure \ref{fig:2}.

\subsection{Resample and Downmix}
Given an incoming 2-channel raw audio with a high sample rate (typically 44.1 KHz), we downsample and downmix it to mono-channel audio.\\

\noindent These operations reduce the memory footprint of the raw audio by an order of magnitude while preserving essential information required to match and identify the relevant audio content.

\subsection{Spectral Representations}

Short-time Fourier Transform (STFT) is applied on the downsampled and downmixed audio to obtain a spectrogram. This spectrogram is further reduced in size by combining individual frequency amplitudes into frequency bands using filter banks. Mel filter bank combines amplitudes from many frequencies into a few mel bands to generate a more compact representation.

\subsection{Transformations}

\subsubsection{Windowing}
\label{sec:window}

In the previous steps, the mel-band amplitudes at a given timestep were extracted, corresponding to an audio segment. A sliding window of a fixed number of timesteps is then applied to generate a mel-spectrogram, as illustrated in Figure \ref{fig:1}. In this representation, each element (X, Y) denotes the amplitude for mel-band Y at timestep X. The sliding window has a small stride, resulting in consecutive mel-spectrograms sharing most of the timesteps. 

A small sliding window enables the generation of a large number of fingerprints. Thus, windowing leads to fingerprints changing gradually over time, retaining some degree of temporal similarity. This characteristic allows for a larger matching region for incoming queries, enhancing the accuracy and robustness of the matching system.

\subsubsection{Time averaging}
We create a one-dimensional array from each mel-spectrogram by taking the running-average amplitude of each mel band across all timesteps. The resulting one-dimensional array has the same number of elements as the mel bands in the original mel-spectrogram.

\subsubsection{Standardization}
The time-averaged mel-band array performs well in terms of match accuracy on clean audio (without much noise). However, for audio received in a production setting, which tends to be noisy, the amplitudes vary significantly when compared to the clean audio. To increase the robustness of the fingerprints and minimize the impact of noise, the mel-band array is standardized to bring the amplitude values within similar ranges and similar kinds of distributions.

\subsubsection{Augmenting with amplitude-delta between consecutive frequency bands}
It is observed that differences in amplitudes between consecutive frequency bands, in addition to the existing input, have a positive impact on match accuracy.

To incorporate this into the fingerprint generation pipeline, the difference between frequency band amplitudes in the time-averaged mel-band array is calculated. It is standardized separately to improve robustness to noise and then appended to the standardized time-averaged mel-band array, giving a higher dimensional array.

\subsubsection{Typecasting}
The values of the spectrogram are in a range such that they can be downcasted from 32-bit floats to 16-bit floats without significant loss in information. This further reduced fingerprint size by 50\%, without affecting match accuracy.

\subsubsection{PCA}
We use Principal Component Analysis (PCA) to further reduce the dimensionality of the fingerprint to 32 dimensions—more compression results in a significant decrease in both the explained variance and match accuracy.

Therefore, this 32-dimensional array is the final output of our proposed fingerprinting algorithm.

\section{Experimental methodology}
\label{sec:exp_method}

\subsection{Fingerprint benchmark}
\label{sec:fp_comp}
For benchmarking our proposed solution, we compare it with a min-hash based audio fingerprint developed by Baluja \& Covell (2006). The min-hash is derived through a sequence of steps, which involve obtaining the bark spectrogram of the audio, selecting the top haar wavelets from the spectrogram's associated windows, and converting the wavelets into two bits by retaining only the magnitude sign. Finally, a min-hash procedure reduces the resulting bit-vector to a 72-dimensional fingerprint. Typically, min-hash fingerprints use a Locality Sensitive Hashing (LSH) for retrieval \cite{Baluja, Wang, Gionis}.
This system of min-hash fingerprinting and LSH retrieval has some limitations.
\begin{itemize}
\item Temporal correlation: The min-hash fingerprint is obtained from a sequence of transformations designed to make it discriminative and robust to various noises. Despite these benefits, the min-hash method exhibits a reduced level of descriptive power and temporal correlation. In particular, for the min-hash, consecutive fingerprints derived from highly overlapping audio segments differ widely in distance.

\item GPU-compatibility: The hamming distance used to compare two min-hash fingerprints is not fully parallelized on GPU in popular ANN libraries, requiring the retrieval algorithm to run on the CPU. 

\item LSH has some drawbacks regarding memory overhead for acceptable results and retrieval speed \cite{Cormode}. These drawbacks are addressed by recent tree-based and graph-based ANN algorithms that scale much better to million and billion-scale vector databases \cite{Johnson, Malkov, Chen, ann_bench}.\\
\end{itemize}

The proposed fingerprint addresses the above limitations and is obtained after a comprehensive exploration of parameters such as the choice of spectral bands, number of bands, robust and discriminative transformations, and fingerprint dimensions, among others. It is optimized for the following properties:
\begin{itemize}
\item Robustness to various types of noise
\item Compatibility with GPU-based search for enhanced retrieval speed
\item High temporal correlation between consecutive fingerprints for improved matching accuracy
\item Sparsity in databases to reduce memory consumption and computational load while increasing retrieval speed
\end{itemize}
In the subsequent section, we describe the experimental setup, terminology, and dataset used.

\subsection{Experimental setup}
\label{sec:exp_setup}
Table~\ref{tab:terminology_table} lists the terminology used in the experimental results.

\subsubsection{Reference database}
\begin{itemize}
\item The reference database utilized in this work consists of approximately thirty million fingerprints extracted from twenty thousand contents. 
\item The database is obtained separately for the min-hash and our proposed fingerprint and is used to create the corresponding index, discussed in the next section.
\end{itemize}

\subsubsection{Index}
\begin{itemize}
\item The properties of the min-hash and proposed indexes are presented in Table~\ref{tab:index_properties}
\item The index created for our proposed algorithm stores significantly fewer fingerprints. Our proposed index is constructed on a skip five database, whereby only one in six consecutive fingerprints are included. Conversely, all fingerprints are stored for the min-hash index from a skip zero database.
\end{itemize}
It is important to highlight that the index parameters for the min-hash and our proposed fingerprint were separately fine-tuned, taking into account the trade-off between match speed, accuracy, index memory size, and build time. The \textbf{optimal index algorithm and parameter settings were selected} to satisfy these criteria for both fingerprinting techniques. The accuracy and speed results presented below are based on the combination of min-hash \& tuned binary CPU-index vs. proposed fingerprint \& tuned L2 GPU-index, even when not explicitly stated. Table~\ref{tab:index_properties} outlines the key differences in the fine-tuned indexes.

\begin{table}[H]
  \caption{Terminology}
  \label{tab:terminology_table}
  \centering
  \begin{tabular}{>{\RaggedRight}p{0.24\linewidth}p{0.67\linewidth}}
    \toprule
    \textbf{Term}      & \textbf{Definition}                \\
    \midrule
    Content                             & An audio clip
    \\
    \hline
    Reference Database (DB)             & Collection of contents used for searching against \\\hline
    Fingerprint (FP)                    & A unique signature for a small segment of content, used to match that content in a DB \\\hline
    Proposed fingerprint                & Our contribution of a robust, GPU-compatible fingerprint in the L2 space with a high temporal correlation \\\hline
    Min-hash                            & The benchmark fingerprint used for comparison against our proposed algorithm \\\hline
    Index                               & Collection of fingerprints corresponding to the contents of a DB and the associated ANN search algorithm \\\hline
    IVF                            & Inverted File index (IVF) is an ANN search algorithm \\\hline
    IVFHNSW                             & A composite index built using the IVF and Hierarchical Navigable Small World (HNSW) search algorithms \\\hline
    Skip Rate                           & The number of consecutive fingerprints skipped for every FP added to an index. It is a metric for sparsity. E.g., A skip 1 DB stores half the fingerprints \\\hline
    Query set                           & The set of FPs used to find a content match in the reference DB using the index \\\hline

    Accuracy                            & Percentage of audio clips for which we correctly identify the content \\\hline
    Fingerprint processing speed (FPS)  & The number of fingerprints searched per second, using an index. A metric for the retrieval speed \\\hline

    Index build time                    & The time taken to train the index from the fingerprints in a DB \\\hline
    Index size                          & The memory size of the index \\\hline
    
    Real setting                        & The production environment where the ACR solution is deployed on a TV \\
    \bottomrule
  \end{tabular}
\end{table}

\begin{table}[H]
  \caption{ANN index properties for the min-hash and our proposed fingerprint}
  \label{tab:index_properties}
  \centering
  \begin{tabular}{>{\RaggedRight}p{0.4\linewidth}p{0.23\linewidth}p{0.2\linewidth}}
    \toprule
    \textbf{Index properties}      & \textbf{Min-hash} & \textbf{Proposed fingerprint}                \\
    \midrule
    Distance metric & Hamming & L2 \\\hline
    ANN index algorithm & Binary IVFHNSW & L2 IVF \\\hline
    Skip in the reference DB (sparsity) & 0 & 5 \\\hline
    Number of fingerprints in the index & $\sim$30 million & $\sim$5 million \\\hline
    GPU compatible & No & Yes \\\hline
    Processor specs & 28 cores of Intel Xeon E5-2697 v3 & one Nvidia 3090 GPU \\
    \bottomrule
  \end{tabular}
\end{table}

\subsubsection{Query set}
For accuracy on artificial noises:
\begin{itemize}
\item One-second segments were taken from 1000 audio contents, and various noise degradations were applied.
\item Our proposed fingerprints and the min-hashes were obtained from the degraded audio, and matching was performed for each one-second segment using both fingerprinting techniques.
\end{itemize}
For accuracy on real noises:
\begin{itemize}
\item Over thirty hours of raw audio were collected from a TV playing multiple channels.
\item This raw audio was divided into five datasets, denoted as Dataset-1 through Dataset-5. For each dataset, both audio fingerprints were obtained.
\item In each dataset, 1.25-second segments of audio fingerprints were taken to find a match for that segment.
\end{itemize}

\subsubsection{Finding a content match}
\label{sec:post_process}
\begin{itemize}
\item The query set was searched using the index to find a content match in the reference database.
\item Post-processing techniques were applied to the fingerprint matches returned by the index to declare a match with high confidence and remove any false matches. In particular,
    \begin{itemize}
    \item An audio segment should have a minimum number of fingerprints matched to a content, i.e., a majority count threshold.
    \item The matched fingerprints to a content should be ordered in time, meaning that a segment of query fingerprints in time should map to roughly consecutive matches in the content.
    \end{itemize}
\end{itemize}

\section{Experimental results}
\label{sec:exp_results}
This section evaluates our proposed fingerprint by considering different properties, which were stated in section~\ref{sec:fp_comp}. To establish the effectiveness of our proposed method, the min-hash fingerprint was taken as a benchmark for comparison (section~\ref{sec:fp_comp}). 

\subsection{Robustness to noise}

\subsubsection{Artificial noises}
Audio fingerprinting techniques must be resilient to various distortions that may occur in real settings (refer to Table~\ref{tab:terminology_table}). Often these distortions are not known beforehand, making it challenging to develop effective fingerprinting algorithms. Consequently, research in audio fingerprinting aims to evaluate the robustness through various artificial noise degradations \cite{Haitsma}. Such evaluations allow for comparisons between different fingerprinting algorithms to identify the superior approach. The noises we experimented with are listed in Table~\ref{tab:artificial_description}. Section~\ref{sec:exp_setup} has the information on the index and query set used for this experiment. 

From Table~\ref{tab:artificial_results}, our proposed fingerprint demonstrates greater or comparable accuracy to the min-hash approach across most artificial noises, with the exception of equalization, frequency masking, and preemphasis. This drop in accuracy may be attributed to the method of normalization employed in our proposed fingerprint. While these three noises selectively distort specific frequency bands, the normalization procedure is applied uniformly across all bands without considering individual weighting.

\begin{table*}[!htbp]
  \caption{Artificial noises used to check the robustness of our proposed fingerprint}
  \label{tab:artificial_description}
  \centering
  \begin{tabular}{>{\RaggedRight}p{0.18\linewidth}p{0.4\linewidth}p{0.2\linewidth}}
    \toprule
    \textbf{Noise}      & \textbf{Description} & \textbf{Parameters tested(X)}                \\
    \midrule
    Frequency masking & Randomly mask X frequency bands at the spectrogram level of 40 Hz each & 5, 10 and 20 bands \\
    Clipping distortion & Audio amplitudes at both the bottom and top X/2th percentile are clipped & 2\%, 10\%, 20\%, 40\% \\
    Equalisation & Adjusts the volume of certain frequency bands by X decibels & 3 dB and 6 dB \\
    Gaussian noise & Add Gaussian noise of fixed amplitude X to the audio signal & 0.01 and 0.02 \\
    Lossy noise & Randomly replace X\% of audio amplitude values with zero & 5\% and 10\% \\
    Shifted noise & Shift the FFT frame boundary by X samples w.r.t the boundary used in the database & 45 and 90 frames \\
    Composite noise & Combination of (Lossy, Gaussian, Shifted) noise & (5\%, 0.01, 45), (10\%, 0.02, 90), (10\%, 0.02, random) \\
    Loudness normalization & Apply a constant gain to match a specific loudness of X LUFS (Loudness Units relative to Full Scale) & -14 and -24 LUFS \\
    Preemphasis & On a signal $\alpha$, use the first order filter: y(t) = $\alpha$(t) - X$\alpha$(t-1) & 0.9 \\
    Time stretch & Speed up or slow down the signal by a factor of X without changing the pitch & 0.9, 0.96, 1.04 and 1.1 \\
    Volume & Increases or decreases the volume of the signal by X decibels & -6 dB and 6 dB \\
    Transcoding & WAV to MP3 conversion at a fixed bitrate of X & 32 and 128 bitrate \\
    \bottomrule
  \end{tabular}
\end{table*}

\begin{table}[H]
  \caption{Accuracy for different kinds of artificial noises}
  \label{tab:artificial_results}
  \centering
  \begin{tabular}{>{\RaggedRight}p{0.5\linewidth}p{0.18\linewidth}p{0.19\linewidth}}
    \toprule
    \textbf{Noise} & \textbf{Min-hash accuracy (\%)} & \textbf{Proposed fingerprint accuracy (\%)} \\
    \midrule
    freq\_mask\_5 & \textbf{99.1} & 88.4 \\
    freq\_mask\_10 & \textbf{95.5} & 72.9 \\
    freq\_mask\_20 & \textbf{79.9} & 46.9 \\
    clipping\_distortion\_2 & 100 & 100 \\
    clipping\_distortion\_10 & 98.2 & \textbf{98.6} \\
    clipping\_distortion\_20 & 93.5 & \textbf{95} \\
    clipping\_distortion\_40 & 74.3 & \textbf{83.1} \\
    equalisation\_3 & \textbf{99.4} & 98 \\
    equalisation\_6 & \textbf{92.6} & 77.3 \\
    gaussian\_noise\_0.01 & 99 & \textbf{99.2} \\
    gaussian\_noise\_0.02 & 95.5 & \textbf{96.8} \\
    lossy\_5\_perc & 99.7 & \textbf{100} \\
    lossy\_10\_perc & 98.9 & \textbf{99.7} \\
    shifted\_45 & 95.4 & \textbf{100} \\
    shifted\_90 & 86.8 & \textbf{100} \\
    composite(5\%, 0.01, 45) & 93 & \textbf{98.9} \\
    composite(10\%, 0.02, 90) & 79.5 & \textbf{94.1} \\
    composite(10\%, 0.02, random) & 85.9 & \textbf{93.1} \\
    loudness\_norm\_-14 & 100 & 100 \\
    loudness\_norm\_-24 & 100 & 100 \\
    preemphasis\_0.9 & \textbf{40.7} & 31.2 \\
    volume\_-6db & 100 & 100 \\
    volume\_6db & 100 & 100 \\
    wav\_to\_mp3\_fixed\_br\_128 & 100 & 100 \\
    wav\_to\_mp3\_fixed\_br\_32 & 100 & 100 \\
    \bottomrule
  \end{tabular}
\end{table}

\noindent Nonetheless, this decline in accuracy is not a significant concern, given that the severity of these distortions is unlikely to occur in the real setting. This is based on the empirical results discussed in section~\ref{sec:real}. It is worth noting that our proposed fingerprints exhibit superior performance in the presence of shifted noise, owing to its high temporal correlation. This is particularly relevant as shifted noise is expected to occur in the real setting, given the uncertainty regarding the starting point of audio fingerprinting for any stream of content played on a television.

\subsubsection{Real noises}
\label{sec:real}
ACR is performed using audio fingerprints obtained from television systems. In this real setting, noise can be introduced at various stages, and the exact nature of this real noise is unknown. Thus, it is crucial to replicate the real setting to evaluate the robustness of our proposed fingerprints to such distortions.

\begin{table}[H]
  \caption{Accuracy on datasets collected in the real setting}
  \label{tab:real_results}
  \centering
  \begin{tabular}{>{\RaggedRight}p{0.3\linewidth}p{0.2\linewidth}p{0.23\linewidth}}
    \toprule
    \textbf{Dataset name} & \textbf{Min-hash accuracy (\%)} & \textbf{Proposed fingerprint accuracy (\%)} \\
    \midrule
    Dataset-1 & 86.88 & \textbf{90.12} \\
    Dataset-2 & 88.7 & \textbf{91.53} \\
    Dataset-3 & \textbf{88.14} & 87.84 \\
    Dataset-4 & 88.37 & \textbf{90.48 }\\
    Dataset-5 & 88.26 & \textbf{90.17} \\
    \bottomrule
  \end{tabular}
\end{table}
\noindent Table~\ref{tab:real_results} presents the accuracy of collected real datasets, showing that our proposed fingerprint is as good as or better than the min-hash. Details of the experimental setup, including the index and query set used, are provided in section~\ref{sec:exp_setup}.

\subsection{Retrieval speed}
In this section, we provide the retrieval speeds for the indexes that achieve the above accuracies, as presented in Table~\ref{tab:retrieval_results}.

\begin{table}[H]
  \caption{ANN index retrieval speeds for the min-hash, and our proposed fingerprint}
  \label{tab:retrieval_results}
  \centering
  \begin{tabular}{>{\RaggedRight}p{0.4\linewidth}p{0.2\linewidth}p{0.2\linewidth}}
    \toprule
    \textbf{Index properties} & \textbf{Min-hash CPU index} & \textbf{Proposed fingerprint GPU index} \\
    \midrule
    Index type & Binary IVFHNSW & L2 IVF \\\hline
    Skip in the reference database (sparsity) & Skip 0 & Skip 5 \\\hline
    Retrieval speed in FPS & 150k & \textbf{4.8M} \\\hline
    Index build time on CPU & 400 mins & 1.2 mins \\\hline
    Index size & 2.3 GB & 0.65 GB \\
    \bottomrule
  \end{tabular}
\end{table}

\noindent Table~\ref{tab:retrieval_results} shows that our proposed fingerprint, when combined with a tuned GPU index is approximately \textbf{30 times faster} in retrieval speed. However, the relative speed-up depends on various factors, such as the model of CPU/GPU used and the number of cores and GPUs used. The gain in speed is due to a combination of factors that favor our proposed fingerprint.
\begin{itemize}
\item It is compatible with a GPU-based search index, which provides faster retrieval speed.
\item It explores a smaller fraction of the search space due to the use of more approximate index parameters. 
\item It has a smaller fingerprint dimension of 32, whereas min-hash has a dimension of 72. Thus, the distance calculation between two fingerprints requires fewer comparisons for our proposed approach.
\item It exhibits a higher temporal correlation, reducing the number of fingerprints that need to be stored in the search index. This is observed by the "Skip in the reference database (sparsity)" field of Table~\ref{tab:retrieval_results}, where only 1 in 6 fingerprints for audio content are stored for our proposed approach. In contrast, all fingerprints are stored for the min-hash index.
\end{itemize}

\subsection{Temporal correlation \& sparsity}

This section investigates the temporal correlation and sparsity properties of our proposed fingerprint. These properties are interrelated, as a fingerprinting technique with high temporal correlation can be used to build a sparser database. When a high temporal correlation fingerprint is added to a DB, we can skip the fingerprints in its vicinity since it already encompasses the information required to detect that audio region. Higher temporal correlation enables us to retain detection information over a larger neighborhood around the fingerprint added to the DB (and skip the remaining fingerprints). This logic for controlling DB sparsity is governed by the skip rate, as described in Table~\ref{tab:terminology_table}.

We present that our proposed approach exhibits a higher temporal correlation than the min-hash. This implies that for our proposed fingerprint, consecutive fingerprints derived from highly overlapping audio regions exhibit slower changes in distance when compared to the min-hash. We perform several experiments to understand and contrast these properties for both fingerprinting techniques.

\subsubsection{Pairwise distances between fingerprints}
\begin{figure}
    \centering
    \includegraphics[width=\linewidth]{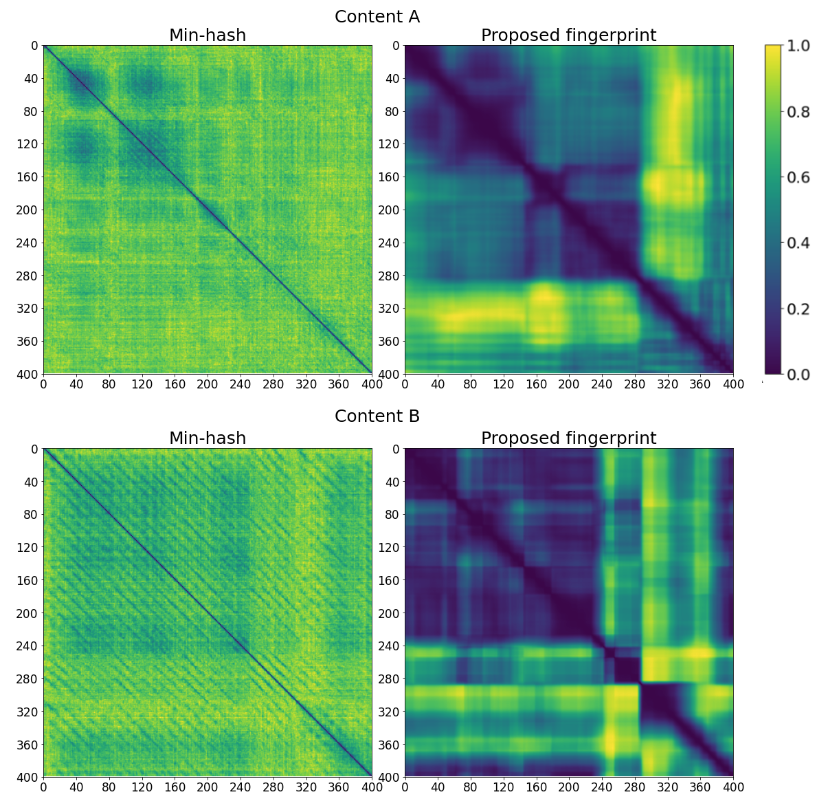}
    \caption{Pairwise distances between min-hash (left) and proposed fingerprint (right) for two contents.}
    \label{fig:3}
\end{figure}

In this experiment, the following steps are performed:
\begin{itemize}
    \item Obtain the min-hash and proposed fingerprint for the audio content
    \item Compute all pairwise distances for both fingerprints.
        \begin{itemize}
        \item Hamming distance was used for the min-hash,
        \item L2 distance for our proposed fingerprint
        \end{itemize}
    \item Min-max normalize the distance matrix to have values between [0, 1]
    \item Visualize the distance matrices for the two fingerprinting techniques, as demonstrated in Figure~\ref{fig:3}. 
\end{itemize}
The pairwise distances between fingerprints within a content determine the similarity between fingerprints occurring from nearby audio regions compared to those from distant audio regions. From Figure~\ref{fig:3}., we have the following observations:
\begin{itemize}
    \item For the min-hash, the distance between fingerprints is low near the diagonal but increases abruptly as we move away from the diagonal, i.e., with increasing gaps between the fingerprints.
    \item In contrast, the distance increases gradually for our proposed fingerprint as we move away from the diagonal. Therefore, consecutive fingerprints are more similar to each other, indicating a higher temporal correlation.
\end{itemize}

\subsubsection{Effect on accuracy with an increase in skip}
The following setup was employed in this experiment:
\begin{itemize}
\item Reference DB: same as the one described in section~\ref{sec:exp_setup}. From this, we construct different sparse DBs based on the skip specified in Table~\ref{tab:skip_results}.
\item Index: An exhaustive search index is used.
\item Query data: Dataset-1.
\end{itemize}

\noindent It should be noted that by utilizing an exhaustive search index, we remove the impact of the index on the accuracy observed in Table~\ref{tab:skip_results}. Instead, the accuracy is attributable to the fingerprinting technique and the skip metric.
 
The results presented in Table~\ref{tab:skip_results} indicate that as the sparsity or skip increases, there is a significant decrease in accuracy for the min-hash. This substantial reduction, however, is not observed for our proposed fingerprint, as the fingerprints retained in the index include the necessary information to identify the audio regions covered by the skipped fingerprints.

\begin{table}[!htbp]
  \caption{Exhaustive index accuracy with an increase in skip rate}
  \label{tab:skip_results}
  \centering
  \begin{tabular}{>{\RaggedRight}p{0.25\linewidth}p{0.2\linewidth}p{0.18\linewidth}p{0.2\linewidth}}
    \toprule
    \textbf{Skip in the reference DB (sparsity)} & \textbf{Number of fingerprints} & \textbf{Min-hash accuracy} & \textbf{proposed fingerprint accuracy} \\
    \midrule
    0 & 30 million & 93.53 & 91.03 \\
    1 & 15 million & 91.65 & 90.9 \\
    3 & 7.5 million & 88.8 & 90.56 \\
    5 & 5 million & 85.46 & 89.96 \\
    7 & 3.75 million & 75.85 & 90.23 \\
    \bottomrule
  \end{tabular}
\end{table}
\section{Conclusion}
\label{sec:conc}
This study presented a novel audio fingerprinting system that employed multiple signal processing techniques to obtain a compact fingerprint for audio. The fingerprint generation process is lightweight and can run on limited computational resources. Our experiments demonstrate that our proposed fingerprint is robust to various forms of artificial noise and noises observed in industry-scale datasets where ACR is performed on a TV. Additionally, we have shown that its robustness is comparable to a min-hash based approach. The compatibility of our proposed fingerprint with GPU search indexes and its high temporal correlation has resulted in a highly scalable retrieval system that is significantly more efficient and faster than the min-hash. The future direction of this study is to enhance our proposed fingerprint's robustness to more adversarial noises and evaluate it on other audio fingerprinting applications.

\end{document}